\documentclass[12pt]{iopart}
\usepackage{graphicx}

\begin{document}

\title{Anti-Newtonian dynamics and  self-induced Bloch oscillations of correlated particles}

\author{Stefano Longhi }

\address{Dipartimento di Fisica, Politecnico di Milano and Istituto di Fotonica e Nanotecnologie del Consiglio Nazionale delle Ricerche, Piazza L. da Vinci
32, I-20133 Milano, Italy}
\ead{longhi@fisi.polimi.it}
\begin{abstract}
 We predict that two correlated particles  hopping on a one-dimensional Hubbard lattice  can show transient self-acceleration and self-induced Bloch oscillations as a result of anti-Newtonian dynamics. Self-propulsion occurs for two particles with opposite effective mass on the lattice and requires long-range particle interaction. A photonic simulator of the two-particle Hubbard model with controllable long-range interaction, where self-propulsion can be observed, is discussed.
\end{abstract}



\maketitle


\section{Introduction}
Physical theories have always rejected the possibility of existence of negative mass,  and very few studies have examined such a hypothetical possibility, usually in the context of gravitational interaction \cite{P1,P2,P3,P4}. Negative mass would lead to some counter-intuitive form of motion and to anti-Newtonian dynamics. One example is provided by self-propulsion, a paradoxical effect which was
suggested by Forward and Millis in the 1990's  \cite{P2,P3}:  an object with negative mass interacting with an object with an equal but positive mass would result in an unlimited amount of unidirectional acceleration of the objects together,  without the need to supply energy or reaction mass.   An alternative way of formulating the negative mass problem is to assume positive mass but anti-Newtonian forces, i.e. forces that break action-reaction symmetry \cite{P5a,P5b}.  While physical particles have always a positive mass and anti-Newtonian models remain fictional theories, it is known that in many physical systems, such as in semiconductors, quantum dot arrays, photonic crystals, cold atoms in optical lattices,  etc.  quasi-particles  may exhibit an effective negative mass. This makes anti-Newtonian dynamics a physically realizable process for quasi-particles.  Despite such a simple observation, anti-Newtonian dynamics for quasi particles has been overlooked by the scientific community.  In Ref. \cite{Batz}, Batz and Peschel suggested an optical realization of self-propulsion for optical pulses in a photonic crystal fiber  exploiting opposite signs of group velocity  dispersion. The  experimental demonstration of this phenomenon has been very recently reported in Ref.\cite{Peschela} by Wimmer and collaborators (see also \cite{Peschelb}) using a nonlinear optical mesh lattice. In such works, particle interaction was mimicked by the Kerr nonlinearity of the fiber.\par
 An important physical model where anti-Newtonian dynamics could be observed is the Hubbard model. The conceptually simplest case is that of two (or few) interacting particles hopping on a one-dimensional lattice, which has been extensively investigated by many authors. In spite of its simplicity, the few-particle Hubbard model contains a rich physics. In ordered lattices, important physical phenomena include the formation of particle bound states and correlated tunneling \cite{cor1,cor2},  robust bound states in the continuum \cite{BICa,BICb}, three-body bound states \cite{Valientea,Valienteb,Valientec}, resonantly enhanced co-tunneling and particle dissociation \cite{Kola,Kolb}, fractional Bloch oscillations \cite{frac}, and correlated Klein tunneling \cite{Klein}, to mention a few. In disordered lattices, great attention has been devoted to study the role of particle interaction (both short- and long-range) on  the localization properties of two correlated electrons \cite{d1,d2a,d2b,d20a,d20b,d2bis,d3a,d3b,d3c,d4}. In two seminal papers \cite{d2a,d2b}, Shepelyansky and Imry found that Hubbard interaction for two particles in the Anderson model can enhance the localization length in comparison with the independent particles. In other cases, such as  in quasi-periodic systems,interaction may generate strongly localized two-particle eigenstates \cite{d5}.
 Such results raised a lively
and controversial debate on the role of interaction and disorder on localization, in particular regarding the functional dependenceof  the localization length on the interaction strength and the disorder \cite{d3a,d3b,d3c}. In spite of such a large amount of studies on two-particle Hubbard systems in either ordered or disordered lattices, the occurrence of anti-Newtonian dynamics and related effects of correlated particles has been so far overlooked.
 \par
 In this work we predict that anti-Newtonian dynamics can arise for certain states of two correlated particles with long-range interaction hopping on an ordered lattice. Like in Refs.\cite{Batz,Peschela}, we exploit the band structure of the lattice to realize effective positive and negative mass.
 An important physical effect of anti-Newtonian dynamics in the Hubbard model, which was not disclosed in Refs.\cite{Batz,Peschela}, is the appearance of transient self-induced oscillations,  which resemble the famous Bloch oscillations \cite{Blocha,Blochb,Blochc,Calla}. Remarkably, Bloch oscillations  in the anti-Newtonian regime are self-induced, i.e. do not require any external force, and are thus very distinct than ordinary Bloch oscillations found so far for correlated particles  \cite{notea,noteb,notec,noted,notee,notef}.\par 
 
 \section{The model}
Let us consider two particles hopping on a one-dimensional lattice in the presence of long-range interaction, either attractive or repulsive; see Fig.1(a). The two particles can be either two electrons with opposite spins in a crystalline potential or two 
atoms trapped in an optical lattice. In the former case, the particles are distinguishable and the hopping dynamics is described by the extended Fermi-Hubbard model (EHM) with Hamiltonian \cite{EHMa,EHMb,EHMc,EHMd}
 \begin{figure}
\includegraphics[scale=0.4]{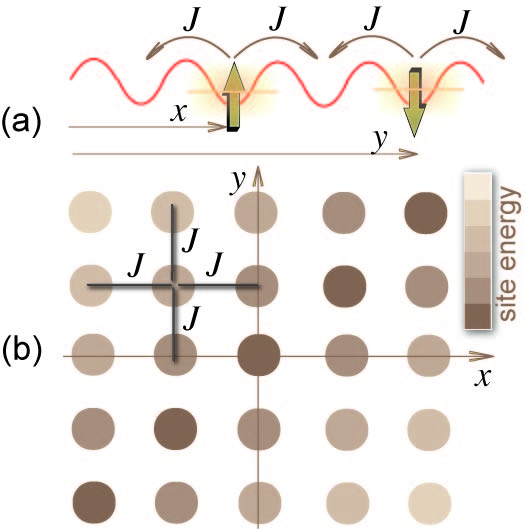}
\caption{(Color online) (a) Schematic of two interacting particles hopping on a one-dimensional lattice (hopping rate $J$), and (b) the equivalent problem of a single particle hopping on a two-dimensional square lattice. Particle interaction renormalizes the site energies of the square lattice. In a photonic simulator of the two-particle Hubbard model the square lattice is realized by evanescently-coupled optical waveguides. }
\end{figure}
\begin{eqnarray}
\hat{H} & = & - J \sum_{l, \sigma= \uparrow, \downarrow} \hat{a}^{\dag}_{l,\sigma} \left( \hat{a}_{l-1,\sigma}+\hat{a}_{l+1,\sigma}  \right) +U \sum_{l} \hat{n}_{l,\uparrow} \hat{n}_{l,\downarrow} \nonumber \\
&+ & \frac{1}{2} \sum_{l \neq k} V(|l-k|) \hat{n}_{l} \hat{n}_{k}
\end{eqnarray}
where $\hat{a}^{\dag}_{l,\sigma}$ are $\hat{a}_{l,\sigma}$ are the fermionic creation and annihilation operators of electrons with spin $\sigma= \uparrow, \downarrow$ at lattice sites $l=0, \pm 1, \pm 2,...$, $J$ is the hopping rate between adjacent sites,  $U$ and $V(|l-k|)$ describe on-site and long-range Coulomb repulsion, respectively, $\hat{n}_{l,\sigma}=\hat{a}^{\dag}_{l,\sigma} \hat{a}_{l,\sigma}$, $\hat{n}_{l}=\hat{n}_{l,\uparrow}+\hat{n}_{l,\downarrow}$ is the particle number operator, and  we assumed $\hbar=1$.  The
EHM is a prototype model in condensed matter theory,
which exhibits a rich phase diagram \cite{EHMa,EHMb,EHMc,EHMd}. In solid-state systems, nonlocal interaction arises from  Coulomb repulsion of electrons in adjacent sites due to nonperfect screening of electronic charges.  Hamiltonian (1) also describes  fermionic ultracold atoms or molecules
with magnetic or electric dipole-dipole interactions in optical lattices  \cite{DMa,DMb,DMc,DMd}. In such physical systems, the non-local interaction energy $V$ shows rather generally a power-law decay with lattice site distance $|l-k|$. Quantum simulators of the Hubbard model with long-range Coulomb interaction using acoustic surface waves have been also suggested \cite{quantumsimulator}. 
In the two-particle sector of Fock space, the amplitude probability $\psi=\psi(\mathbf{r},t)$ to find the electron with spin $\uparrow$ at lattice site $x$ and the electron with spin $\downarrow$ at lattice site $y$ evolves according to the Schr\"{o}dinger equation $ i \partial_t \psi= \hat{H}_0( \mathbf{r}, \hat{\mathbf{p}}) \psi$ with Hamiltonian
\begin{equation}
\hat{H}_0( \mathbf{r}, \hat{\mathbf{p}})=-2J \cos (\hat{p}_x)-2J \cos(\hat{p}_y)+W(x,y)
\end{equation}
where we have set $\mathbf{r}=(x,y)$, $\hat{\mathbf{p}}=-i \nabla_{\mathbf{r}}=(-i \partial_x, -i \partial_y)$, $W(x,y)=V(|y-x|)$ for $x \neq y$, $W(0)=U$.
 Note  that, since $\hat{H}_0 \rightarrow - \hat{H}_0$ for $\hat{p}_{x,y} \rightarrow \hat{p}_{x,y}+\pi$ and $W \rightarrow -W$, the particle dynamics is the same for attractive or repulsive interaction; hence we can limit to consider either the repulsive ($W>0$) or attractive ($W<0$) case. The energy spectrum $E$ of the two-particle states is purely continuous and can be exactly determined from $\hat{H}_0( \mathbf{r}, \mathbf{p})$ with the Ansatz $\psi(x,y,t)=f(y-x,K) \exp[iK(x+y)/2-iEt]$, where $K$ is the total quasi-momentum of the two particles. This maps the two-particle Hamiltonian problem into a single particle problem for each value of the total quasi-momentum \cite{Valiente2a,Valiente2b}, namely 
 \begin{eqnarray}
 -2 J \cos (K/2)  [ f(s+1,K)+f(s-1,K) ] \nonumber \\
 + V(|s|)f(s,K) =E(K) f(s,K)
 \end{eqnarray}
 where $s=y-x$.
   \begin{figure}
\includegraphics[scale=0.5]{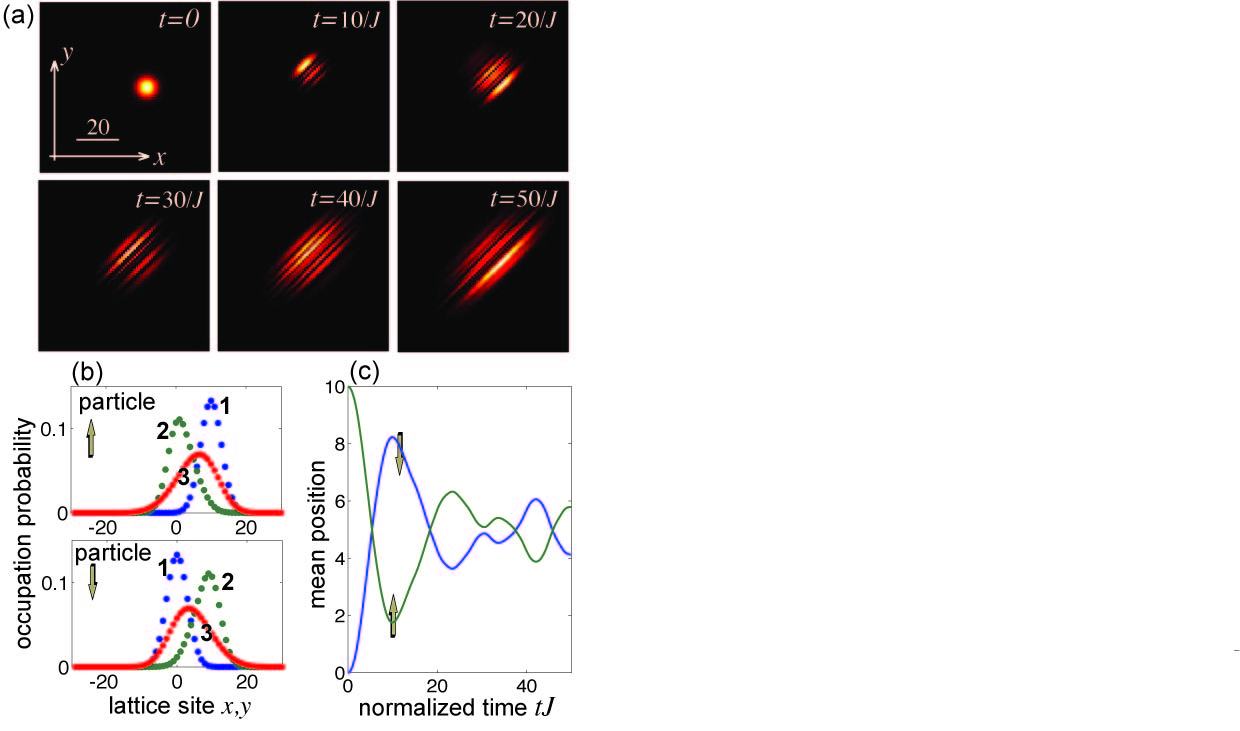}
\caption{(Color online) (a) Evolution of the two-particle site occupation probability $|\psi(x,y,t)|^2$ at a few times $t$ for two initial particles initially at rest with the same mass ($p_x=p_y=0$), corresponding to Newtonian dynamics. Parameter values are given in the text. (b) Detailed behavior of the single particle site occupation probabilities $P_{\uparrow}(x,t)=\sum_y |\psi(x,y,t)|^2$ and 
$P_{\downarrow}(y,t)=\sum_x |\psi(x,y,t)|^2$
at the three times $t=0$ (curve 1), $t=10/J$ (curve 2), and $t=20/J$ (curve 3). (c) Numerically-computed evolution of the mean positions $\langle x(t) \rangle$ and $\langle y(t) \rangle$ for the two particles with spins $\uparrow$ and $\downarrow$.}
\end{figure}
  \begin{figure}
\includegraphics[scale=0.5]{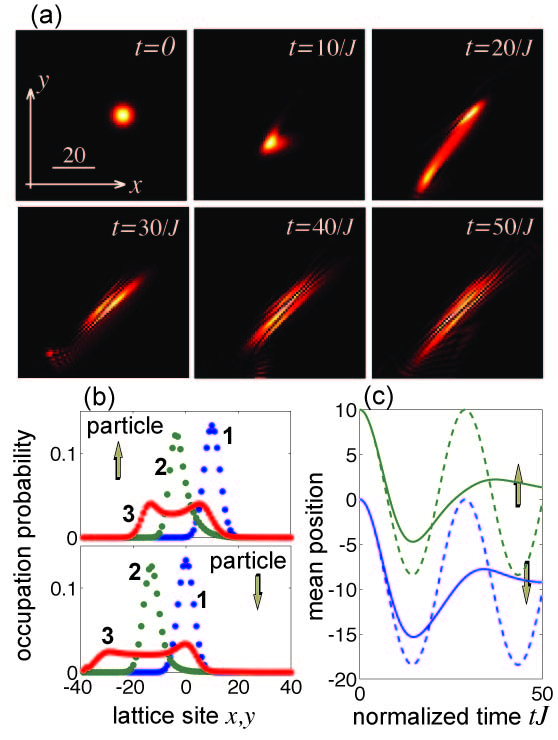}
\caption{(Color online) Same as Fig.2, but for the two particles with opposite mass ($p_x=0$, $p_y= \pi$), corresponding to anti-Newtonian dynamics. In (c) the dashed curves show the oscillatory paths of the two particles due to self-induced Bloch oscillations as predicted by the semiclassical analysis [Eq.(8)].}
\end{figure}
 The solutions to Eq.(3) classify into symmetric (S) and antisymmetric (A) scattered states $f^{(S,A)}(s,K,q)$ with energy $E(K,q)=-4 J \cos (K/2) \cos q$ and asymptotic behavior $f^{(S)}(s,K,q) \sim \cos(q|s|+\delta_S)$ and $f^{(A)}(s,K,q) \sim (s/|s|) \cos(q|s|+\delta_A)$ as $|s| \rightarrow \infty$, where $\delta_{S,A}(q)$ are phase shifts and $q$ the relative quasi momentum of the particles; and a set of symmetric or antisymmetric bound states $f^{(n)}(s,K)$ ($|f| \rightarrow 0$ as $|s| \rightarrow \infty$)  with energy $E_n(K)$ satisfying the constraint $|E_n(K)|>4J | \cos (K/2) |$ \cite{Valiente2a,Valiente2b}. In the original two-particle problem, the eigenfunctions $f^{(S,A)}(s,K,q)$ correspond to unbound particle states, fully delocalized in the lattice, whereas the eigenfunctions  $f^{(n)}(s,K)$   correspond to particle bound states (doublons), which are nevertheless fully delocalized along the lattice. 
 
 \section{Semiclassical analysis and self-induced Bloch oscillations}
 An initial two-particle wave packet can be decomposed as a superposition of scattered and doublon states of $\hat{H}_0$, and its temporal evolution is governed by the interference among such spectral components. For spatially-broad two-particle wave packets, the hopping motion  on the lattice is at best captured by a semiclassical analysis of the Hamiltonian $\hat{H}_0$ \cite{Calla}. The expectation value $\langle A \rangle= \sum_{x,y,t} \psi^*(x,y) \hat{A}(\mathbf{r},\mathbf{p}) \psi(x,y,t)$ 
 of any operator $\hat{A}$ evolves according to the equation $i d \langle A \rangle  /dt=\langle [\hat{A},\hat{H}_0] \rangle$. In particular, the mean values of position $\langle x \rangle$, $\langle y \rangle$ and quasi-momenta $\langle \hat{p}_x \rangle$, $\langle \hat{p}_y \rangle$ of the two particles on the lattice satisfy the (exact) Ehrenfest equations
 \begin{eqnarray}
 \frac{d \langle x \rangle}{dt} & = & 2 J \langle \sin \hat{p}_x \rangle  , \;\;\;\;  \frac{d \langle \hat{p}_x \rangle}{dt}  =  - \langle \frac{\partial W}{ \partial x}\rangle \\ 
 \frac{d \langle y \rangle}{dt} & = & 2 J \langle \sin \hat{p}_y \rangle ,  \;\;\;\;  \frac{d \langle \hat{p}_y \rangle}{dt}  =  - \langle \frac{\partial W}{ \partial y}\rangle 
 \end{eqnarray}
 Note that, since $W(x,y)=V(|y-x|)$ and thus $(\partial W/ \partial x)=-(\partial W/ \partial y)$, one has $d (\langle \hat{p}_x \rangle+\langle \hat{p}_y \rangle )/dt=0$, i.e. the total particle quasi-momentum is conserved. However, the effective masses of the two particles are not conserved owing to the band structure of the lattice, and they can assume positive or negative values, a condition which is necessary to observe particle self-propulsion. This is clearly seen in the semiclassical limit, which is obtained by assuming that (i) the potential $W(x,y)$ varies slowly over one lattice site, (ii) the particles are spatially separated with a localization length over which the interaction $W$ is almost contant, and (iii) the particle quasi-momentum distributions are narrow at around their mean values $\langle \hat{p}_x \rangle$, $\langle \hat{p}_y \rangle$. Under such assumptions, Eqs.(4) and (5) simplify into  
  \begin{eqnarray}
 \frac{d \langle x \rangle}{dt}    \simeq   2 J  \sin \langle \hat{p}_x \rangle  , \;\;\;\;  \frac{d \langle \hat{p}_x \rangle}{dt}  \simeq    \frac{\partial V}{ \partial s} \left( | \langle y \rangle - \langle x  \rangle | \right)  \;\;\;\; \\
 \frac{d \langle y \rangle}{dt}    \simeq   2 J  \sin \langle \hat{p}_y \rangle  , \;\;\;\;  \frac{d \langle \hat{p}_y \rangle}{dt}  \simeq  -  \frac{\partial V}{ \partial s} \left(| \langle y \rangle - \langle x  \rangle | \right). \;\;\;
  \end{eqnarray}
  with $s=y-x$.  The main prediction of the semiclassical equations is that the particle motion strongly depends on the initial particle states, and that for {\it certain} initial states anti-Newtonian dynamics and self-induced Bloch oscillations can be observed. Let us first assume that at initial time the two particles are at rest with $\langle p_x(0) \rangle=\langle p_y(0) \rangle=0$. It then follows that $\langle p_x \rangle=-
\langle p_y \rangle$ at any subsequent time, and thus the center of mass  $(\langle x \rangle +\langle y \rangle)/2 $ of the two particles on the  lattice remains at rest: this is the usual scenario of Newtonian dynamics, where two interacting particles accelerate into opposite directions owing to their mutual attractive or repulsive interaction. Conversely, let us suppose that at initial time the two particles have  quasi-momenta that differ by $\pi$, which corresponds to have equal effective masses in absolute value but of opposite sign. In this case one has $\langle p_x \rangle+
\langle p_y \rangle = \pm \pi$ at any successive time, and thus the mean distance of the particles $\langle y \rangle- \langle x \rangle$ on the lattice does not change in time: this means that the two particles accelerate in the same direction, i.e. self-propulsion is attained.  Noticeably, since the particle distance and thus the force $F=-(\partial V/ \partial s)$ do not change during the motion, from Eqs.(6) and (7) it follows that the quasi-momenta linearly increase with time and  the particles undergo an  oscillatory motion which is analogous to the famous Bloch oscillations \cite{Calla}. Note that the oscillatory motion is not induced by any external force, like in the ordinary Bloch oscillations, rather it is self-sustained by the mutual particle interaction, provided that at initial time the two particles have opposite effective masses. For example, let us assume that the two particles are initially at rest with $\langle p_x \rangle=0$, $\langle p_y \rangle =\pi$, and let us 
assume a long-range interaction with exponential tails of the form $W(s)=U \exp(- \gamma s)$. Indicating by $d$ the initial separation of the two particle wave packets,  the constant force acting on them is given by $F=\gamma U \exp(-\gamma d)$, and the semiclassical equations (6) and (7) predict an oscillatory trajectory of the two particles with period $2 \pi /F$, namely 
\begin{equation}
\langle x(t) \rangle= \langle x(0) \rangle +\frac{2J}{F} \left[ \cos(Ft)-1 \right]
\end{equation}
and $ \langle y(t) \rangle=\langle x(t) \rangle -d$. 
The predictions of the semiclassical analysis are approximate results that hold only transiently; indeed, the spectrum of $\hat{H}_0$ is purely continuous and any wave packet $\psi(x,y,t)$, given by the superposition of scattered and doublonic  states,  broadens and breaks up, with the component of doublonic states remaining localized around $x \sim y$. Hence, after an initial transient the semiclassical analysis fails, and the exact dynamics should be investigated by direct numerical analysis of the two-particle Hamiltonian (2). \par
The previous analysis is rather general and can be applied to  either quantum or classical systems, such as ultracold atoms, optical or acoustical systems, where the underlying dynamics can be effectively described by the EHM. As discussed above, the observation of anti-Newtonian dynamics and self-induced BOs requires long-range interaction and controllable initial states of the two particles, which could be of difficult experimental implementation using ultracold atoms in optical lattices or other quantum systems. On the other hand, transport of classical light in a square lattice of optical waveguides \cite{frac,Koma,Komb} can provide a feasible testbed for the observation of self-induced BOs. As is well-known (see, for instance, \cite{frac,Koma,Komb}), in such an optical system the two-particle EHM is mapped into the equivalent single-particle tight-binding problem (2), in which a single particle hops on a two-dimensional square lattice $(x,y)$;  see Fig.1(b). The lattice is realized by a bi-dimensional square array of evanescently-coupled optical waveguides with a straight axis, and the temporal dynamics of the quantum model is mapped into spatial evolution of the light beam along the array. The long-range interaction potential $W$ introduces site-energy renormalization, mimicked by waveguide propagation constant mismatch along the lattice diagonals \cite{frac}; see Fig.1(b). Gaussian beam excitation of the lattice at a tilted angle from the input facet provides the suitable 
initial condition to observe transient self-propulsion and Bloch oscillations. Such an excitation corresponds to the regime of hyperbolic diffraction,  where the effective diffraction of light along two orthogonal directions has an opposite sign owing to the band structure of the lattice (see, for instance, \cite{Stala,Stalb}). In the optical realization of the long-range Hubbard model shown in Fig.1(b), self-induced Bloch oscillations can be thus viewed as an oscillatory motion of the optical beam that arises from the interplay between hyperbolic diffraction and local index gradient. We note that, while in the optical systems of Refs.\cite{Batz,Peschela} self-propulsion requires nonlinear beam interaction, in our waveguide lattice light transport is linear. Moreover, we predict here a novel phenomenon associated to anti-Newtonian dynamics, namely transient  {\it self-induced} Bloch oscillations.  \par
The predictions of the semiclassical analysis have been checked by numerical simulations of the two-particle EHM [Eq.(2)].  In the simulations, an exponential shape $W(s)=U \exp(-\gamma s)$ for the particle interaction has been assumed, and the attractive case ($U<0$) has been considered. As an initial condition, a Gaussian two-particle wave packet $\psi(x,y,0) \propto \exp[-(x-x_0)^2/w^2-(y-y_0)^2/w^2-i p_x x]$  is assumed with either $p_x=0$ or $p_x=\pi$, corresponding to two particles initially at rest at the distance $d=|y_0-x_0|$  with equal ($p_x=0$) or opposite ($p_x= \pi$) effective mass. As an example, in Figs.2 and 3 we show the numerical results of wave packet propagation and particle trajectories as obtained for $U/J=-6$, $\gamma=1/12$, $w=6$, $d=10$ and for $p_x=0$, i.e. for the two particles with the same mass (Fig.2), and for $p_x=\pi$, i.e for the two particles with opposite effective mass on the lattice (Fig.3). For such parameter values, a large number ($\sim 25$) of doublonic states $f^{(n)}$  are sustained by the Hubbard Hamiltonian at $K=0$. The lattice comprises 80 sites, and the dynamics is observed up to the time $t=50/J$ to avoid particle reflections at the lattice edges. 
As expected,  in the former case (Fig.2) a Newtonian dynamics is observed: the particles attract each other, they accelerate in opposite directions and the center of mass remains at rest. A different behavior is found for  the $p_x=\pi$ case (Fig.3), corresponding to the two particles having the same mass in absolute value but with opposite sign. An anti-Newtonian dynamics is here clearly observed:  the particles accelerate in the same directions and their distance $\langle y \rangle -\langle x \rangle$ does not change in time. The particle trajectories show a damped oscillatory behavior, which deviates from the Bloch-oscillation semiclassical path [Eq.(8)] for times $t > \sim 15/J$ [see Fig.3(c)]. The reason thereof is due to wave packet  broadening, break up and elongation along the diagonal $y=x$, which makes the semiclassical limit invalid after an initial transient. Nevertheless, the signature of self-induced Bloch oscillations and reversal of the acceleration is clearly visible. To get an idea of the physical parameters corresponding to the plots of Figs.2 and 3, let us consider a photonic simulator of the two-particle Hubbard model \cite{frac,Koma,Komb}, see Fig.1(b). For a square lattice of evanescently-coupled optical waveguides manufactured in fused silica by femtosecond laser writing and probed in the red ($\lambda=933$ nm), for a lattice constant $a=13 \; \mu$m  the coupling $J$ turns out to be $J \sim 2.5 \; {\rm cm}^{-1}$ according to the data of Ref.\cite{frac}. An evolution time up to $t=20/J$, which is sufficient to observe self-propulsion and reversal of acceleration, corresponds to a sample  of length $\sim 8$ cm, which is feasible with the current technology. The initial two-particle wave packet conditions of Figs.2 and 3 correspond to the excitation of the lattice with a Gaussian beam of spot size $w=6a=78 \; \mu$m at either normal incidence (to observe Newtonian dynamics, Fig.2) or tilted in the vertical $y$ direction by the Bragg angle $\theta_B= \lambda/(2a) \simeq 1.39^{\rm o}$ (to observed anti-Newtonian dynamics, Fig.3). As a final comment, we note that the previous analysis considered distinguishable particles, however similar results (including anti-Newtonian dynamics and self-induced Bloch oscillations) can be observed for two indistinguishable bosons, where the additional symmetry constraint  $\psi(x,y,t)=\psi(y,x,t)$ of the wave function should be imposed. 
\section{Conclusions}
In this work we have theoretically predicted that anti-Newtonian dynamics can spontaneously occur for  two correlated particles with long-range interaction hopping on one-dimensional lattices. Our results disclose novel interaction-induced transport phenomena in Hubbard models, like self-induced transient Bloch oscillations. Such oscillations do not require any external force and arise from the interlay between Bragg scattering in the lattice and the self-interaction of the two particles. The observation of anti-Newtonian dynamics and self-induced BOs requires long-range particle interaction and the control of the initial state of the system, with the two particle wave packets experiencing opposite effective masses on the lattice. Light transport in a square lattice of evanescently-coupled optical waveguides can provide an experimentally accessible testbed to observe anti-Newtonian dynamics and self-induced BOs. In this optical system, self-induced BOs can be explained as a result of the interplay between hyperbolic diffraction \cite{Stala,Stalb} in  the two-dimensional lattice and a local index gradient, that mimics long-range particle interaction. We note that, as compared to the recent experimental demonstration of diametric drive acceleration reported by Wimmer and collaborators  for optical pulses propagating in a nonlinear optical mesh lattice \cite{Peschela}, the Hubbard model and its optical realization proposed in the present work differ in two important respects: (i) action-reaction symmetry breaking in the two-particle Hubbard model with long-range interaction leads to transient self-induced Bloch oscillations, a dynamical regime which was not disclosed in Ref.\cite{Peschela}, and (ii) the optical setup proposed to observe self-induced Bloch oscillations entails {\it linear} propagation of discretized light, i.e. optical (Kerr) nonlinearity is not required. It is envisaged that  the present results could inspire new approaches in controlling wave dynamics in crystal and optical lattices,  stimulating further investigations on non-Newtonian dynamical regimes in few-particle systems.

\end{document}